\documentclass[9pt,journal]{IEEEtran}

\ifCLASSINFOpdf
  \usepackage[pdftex]{graphicx}
\else
\fi

\usepackage{amsmath,amsfonts,amssymb,bm}
\usepackage{multicol}
\usepackage{multirow}
\usepackage{booktabs}
\usepackage{xcolor}

\DeclareMathOperator*{\argmin}{arg\,min}
\newcommand{\norm}[1]{\left\lVert#1\right\rVert}

\newcommand{\bp}{\bm{p}}
\newcommand{\bb}{\bm{b}}
\newcommand{\bx}{\bm{x}}
\usepackage{paralist}
\newcommand{\MZ}[1] {{#1}} %

\begin{document}
\title{Motion Correction via Locally Linear Embedding for Helical Photon-counting CT}

\author{Mengzhou~Li,
    Chiara Lowe,
    Anthony Butler,
    Phil Butler,
    and~Ge~Wang$^*$,~\IEEEmembership{Fellow,~IEEE}%
    \thanks{M. Li and G. Wang are with the Department
        of Biomedical Engineering, Rensselaer Polytechnic, Troy,
        NY, 12180 USA (wangg6@rpi.edu).}%
    \thanks{C. Lowe, A. Butler and P. Butler are with University of Canterbury and MARS Bioimaging Ltd, Christchurch, New Zealand.}%
}

\maketitle
\thispagestyle{empty} %
\pagestyle{empty}

\begin{abstract}
    X-ray photon-counting detector (PCD) offers low noise, high resolution, and spectral characterization, representing a next generation of CT and enabling new biomedical applications. It is well known that involuntary patient motion may induce image artifacts with conventional CT scanning, and this problem becomes more serious with PCD due to its high detector pitch and extended scan time. Furthermore, PCD often comes with a substantial number of bad pixels, making analytic image reconstruction challenging and ruling out state-of-the-art motion correction methods that are based on analytical reconstruction. In this paper, we extend our previous locally linear embedding (LLE) cone-beam motion correction method to the helical scanning geometry, which is especially desirable given the high cost of large-area PCD. In addition to our adaption of LLE-based parametric searching to helical cone-beam photon-counting CT geometry, we introduce an unreliable-volume mask to improve the motion estimation accuracy and perform incremental updating on gradually refined sampling grids for optimization of both accuracy and efficiency. Our numerical results demonstrate that our method reduces the estimation errors near the two longitudinal ends of the reconstructed volume and overall image quality. The experimental results on clinical photon-counting scans of the patient extremities show significant resolution improvement after motion correction using our method, which reveals subtle fine structures previously hidden under motion blurring and artifacts.
\end{abstract}

\begin{IEEEkeywords}
    Rigid motion, motion estimation, locally linear embedding, helical CT, photon-counting detectors (PCDs).
\end{IEEEkeywords}

\IEEEpeerreviewmaketitle

\section{Introduction}

\IEEEPARstart{M}{otion} induced image artifacts are a long-standing problem in X-ray computed tomography (CT), which compromises many clinical CT scans. For instances, head CT and cardiac CT often suffer from image artifacts due to motions of either rigid structures (such as the head) or nonrigid tissues and organs (such as the heart). In this study, we focus on the rigid motion estimation and associated artifact reduction. The rigid motion of an object can be equivalently treated as the view-dependent transformation of the locations of the source and detector pair. Without proper compensation for the motion, the reconstructed image quality will be degraded by the geometric errors, resulting in blurring, double-edging, and strong streaks. Correction for rigid motion can be performed in an iterative reconstruction framework, which minimizes a loss function by alternatively updating image reconstruction and motion estimation until convergence~\cite{sun2016iterative}. However, this method is slow due to the computational cost \MZ{of a large number of iterative reconstructions needed}. For reconstruction acceleration, the modified FDK~\cite{nuyts2020iterative} and weighted filtered backprojection (WFBP) ~\cite{jang2019head} were developed to accommodate non-regular geometry for motion compensation. Along another direction, a powerful parallel searching algorithm based on locally linear embedding (LLE) was proposed for cone-beam CT~\cite{chen2018general}, which replaces \MZ{sequential} gradient-base optimization steps and offers a significant speed boost aided by GPU.

The emerging X-ray photon-counting detector technology promises to revolutionize the CT field, with the recent milestone that the Siemens photon-counting CT scanner received the FDA approval last year. However, the current photon-counting detector is still far from being perfect~\cite{li2020x,getzin2019non}, which brings challenges to the motion estimation and correction. First, due to the complex manufacturing process, PCDs on the market come with a significant amount of ineffective pixels as a result of manufacturing cost control. Also, a large PCD array often consists of tiled small PCD chips. In the tiling process, gaps are formed to give a sufficient room for electronics. Both the gaps and the ineffective pixels are called bad pixels that fail to record projection values. Often times, the bad pixels are so many that the missing values cannot be satisfactorily addressed using classic interpolation methods. As a result, it is difficult for traditional analytical reconstruction to be applied. Second, given a much higher cost of a PCD than a flat panel detector of the same-area, a helical scan is often performed for an extended longitudinal field of view (FOV). These two issues in this context make the motion estimation and correction a unique problem.  

In this paper, we follow the state-of-the-art LLE-based motion estimation idea~\cite{chen2018general} to address the above challenge. In the next section, we describe a LLE-based method for helical photon-counting CT. In the third section, we report our preliminary results. To conclude the paper, we discuss relevant issues in the last section.

\section{Methodology}
\subsection{LLE-based motion correction}
A rigid motion of a moving object is equivalent to a geometrical misalignment of the source and detector pair around a stationary object. The goal is to minimize the difference between measured projections and re-projected projections of an image volume reconstructed assuming estimated motion parameters. Mathematically, we formulate the optimization problem as follows:
\begin{equation}
    \argmin_{\bp,\bx} \norm{\bb - A(\bp)\bx}^2_2, \label{eq:projLoss}
\end{equation}
where \(A(\bm{p})\) is a system matrix after incorporating a vector motion parameters \(\bp\), \(\bx\) and \(\bb\) denote a reconstructed image and measured projections respectively.
The optimization is often performed by alternatively updating \(\bp\) and \(\bx\). Since the motion parameters for one view is separable from those for other views, the problem can be dealt as a set of sub-problems for each view to update estimated motion parameters:
\begin{equation}
    \bp=[\bp_1,\cdots,\bp_i,\cdots,\bp_N]^T,\;\bp_i = \argmin_{\bp_i} \norm{\bb_i - A_i(\bp_i)\bx}^2_2,
\end{equation}
where \(i=1,2,\cdots, N\) denotes the \(i\)th view out of the \(N\) views in total, and \(\bp_i\) is the motion parameters for the \(i\)th view in the form of \([t_x,t_y,t_z,\theta_x,\theta_y,\theta_z]^T\) characterized by six degrees of freedom, describing the translation and rotation along and around three Cartesian axes respectively.

Different from gradient-based optimization, LLE utilizes a parallel searching strategy by densely sampling a pre-defined parametric range. The basic idea is that if the sampling grid is sufficiently small, the true vector of motion parameter should be so close to its \(K\)-nearest neighbors in the sampling grid and can be expressed as the a linear combination of them such that its corresponding projection measurement can also be represented with the linear combination of the \(K\)-nearest reprojected projections associated with the sampled neighboring parameters and with the same weights, i.e.,
\begin{equation}
    \bp_i^* = \sum_{k=1}^{K} w_{k}\bp_i^{(k)}, \quad \bb_i = \sum_{k=1}^{K} w_{k} A_i(\bp_i^{(k)})\bx, \label{Eq:LLE}
\end{equation}
where \(\sum_{k=1}^K w_k = 1\), and \(\bp_i^{(k)}\) is one of the \(K\) nearest samples for \(\bp_i^*\) on the grid \((\bp_i^{(1)},\cdots,\bp_i^{(j)},\cdots,\bp_i^{(S)})\) with \(S\) samples in total, and \(\bp_i^*\) denotes the ground truth motion parameters. This can be understood in light of the Taylor expansion of \(A_i\) at \(\bp_i^*\):
\begin{equation}
    A_i(\bp_i) \approx A_i(\bp_i^*) + (\bp_i-\bp_i^*)^T \frac{\partial A_i(\bp_i^*)}{\partial\bp_i},
\end{equation}
Hence,
\begin{align*}
    \sum_{k=1}^{K} w_{k} A_i(\bp_i^{(k)}) & = \sum_{k=1}^{K} w_{k} (A_i(\bp_i^*) + (\bp_i^{(k)}-\bp_i^*)^T \frac{\partial A_i(\bp_i^*)}{\partial\bp_i})  \\
                                          & = A_i(\bp_i^*)  +  (\sum_{k=1}^{K} w_{k} (\bp_i^{(k)}-\bp_i^*)^T)\frac{\partial A_i(\bp_i^*)}{\partial\bp_i}
\end{align*}
If we have \(\bp_i^* = \sum_{k=1}^{K} w_{k}\bp_i^{(k)}\), we can easily reach the conclusion:
\begin{equation}
    \sum_{k=1}^{K} w_{k} A_i(\bp_i^{(k)}) = A_i(\bp_i^*),\; \bb_i = \sum_{k=1}^{K} w_{k} A_i(\bp_i^{(k)})\bx.
\end{equation}

\MZ{In practice, sampling a multidimensional grid densely is too expensive. Instead, for each sub-problem we estimate the parameters of \(\bp_i\) individually and sequentially (six stages in total) as an effective approximation. All sub-problems can be simultaneously solved and updated for each stage.}
Specifically, for the \(r\)th parameter \(\bp_{ir}\) in one subproblem to be updated, the key steps are as follows:
\begin{enumerate}
    \item Generate a dense sample grid \(\{\bp_{ir}^{(s)}|s = 1,\cdots, S\}\) for \(\bp_{ir}\);
    \item Calculate the reprojected projection grid \(\{(\bb_i^{(s)}, \bp_{ir}^{(s)}) |s = 1,\cdots, S| \}\) by replacing the \(r\)th parameter of \(\bp_i\) with the sampled values on the above grid;
    \item Find the \(K\) nearest neighbors \(\{(\tilde{\bb}_i^{(k)}, \tilde{\bp}_{ir}^{(k)}) | k = 1,\cdots,K\}\) for \(\bb_i\) from the projection grid in terms of the Euclidean distance;
    \item Optimize the weights \(\bm{w} = [w_1,\cdots,w_K]^T\) for the K neighbors to fit the measurement:
          \begin{equation}
              \bm{w} = \argmin_{\bm{w}} \norm{\bb_i - \sum_{k=1}^K{w_k \tilde{\bb}_i^{(k)}}} \, \text{s.t.} \, \sum_{k=1}^K{w_k} = 1; \label{eq:LLE}
          \end{equation}
    \item Update the estimated parameter as \(\bp_{ir} = \sum_{k=1}^K{w_k\tilde{\bp}_{ir}^{(k)}} .\)
\end{enumerate}
Parallel computation can be implemented on GPU in a view-independent fashion for each degree of freedom, and the whole process can be iterated a few times to reach convergence (please refer to \cite{chen2018general} for more details).

\subsection{Bad pixel masking}
\begin{figure}[!t]
    \centering
    \includegraphics[width=\linewidth]{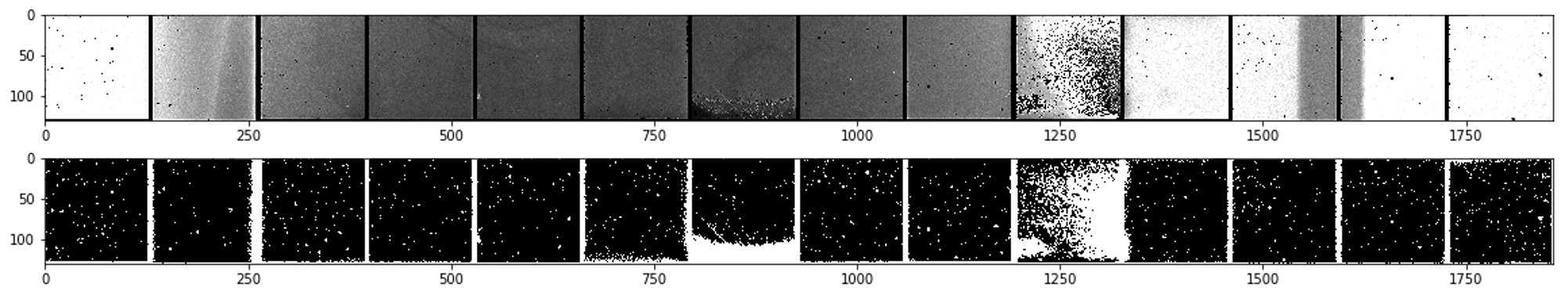}
    \caption{Exemplary projection (top) and its bad pixel mask (bottom).}
    \label{fig:RawPj}
\end{figure}

Most PCDs contain a substantial amount of bad pixels, including real ineffective pixels from the manufacturer, the tile gaps between chips, and dead/bad pixels due to degradation over time. These bad pixels produce unreliable responses and in a much larger number than that of a traditional energy-integrating detector. One exemplary projection with a 14-chip PCD is shown in Fig.~\ref{fig:RawPj}.
Since the number of bad pixels is so significant that a typical interpolation method is inadequate to address the problem without introducing major artifacts. Thus, bad pixel responses will directly reduce the motion estimation accuracy.

Our strategy to avoid the issue is to turn off the bad pixels with a binary mask \MZ{and utilize iterative reconstruction methods to avoid their contribution to the reconstruction and motion estimation}. We use two criteria for detection of bad pixels with a open beam projection data: (1) the temporal mean of the pixel values is a statistically outlier in the group of all pixels; (2) the temporal variance of the pixel is a statistically outlier in the group of all pixels. The mask for the bad pixels will be used to exclude contributions from bad pixels \MZ{during the volume reconstruction and} in the calculation of the fidelity loss in Eq.~\ref{eq:LLE} and facilitate motion estimation and image reconstruction.

\subsection{Unreliable volume masking}\label{sec:masking}
\begin{figure}[!t]
    \centering
    \includegraphics[width=1.6in]{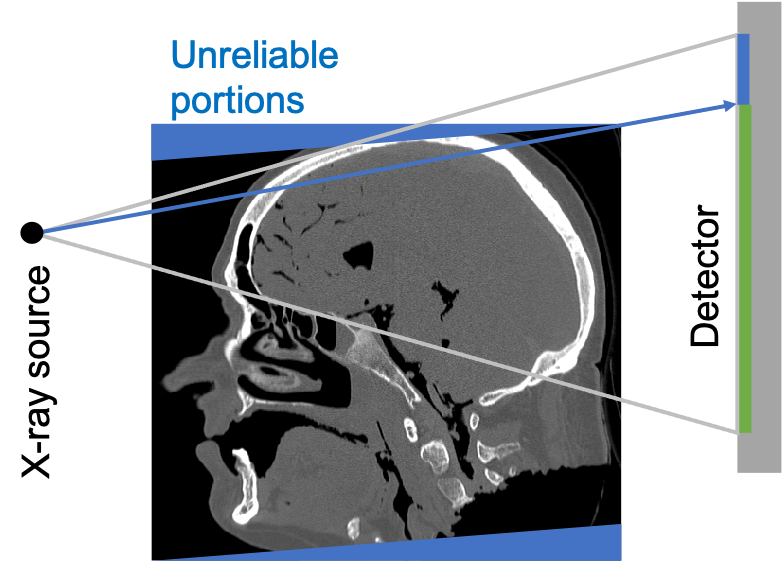}
    \caption{Illustration for the reliable volume masking. Any ray through an unreliable portion (marked in blue on the image volume) is unreliable and contributes to the mask indicating an unreliable area in a projection (marked in blue on the detector) which should be excluded in the loss calculation for motion estimation.}
    \label{fig:ReliablilityMask}
\end{figure}

Due to the axial truncation in the cone-beam geometry/helical scan, there are unreliable portions at two longitudinal ends of a reconstructed volume due to data insufficiency, as illustrated in Fig.~\ref{fig:ReliablilityMask}. 
\MZ{Since the reconstruction of such a region is solely determined by a very limited angular range of projections resulting a severe ill-posed problem, the strong limited angle artifacts will counter-balance the projection loss (Eq.~\ref{eq:projLoss}) originated from misalignment and degrade the motion estimation accuracy.}
The original cone-beam LLE MC method~\cite{chen2018general} does not consider this defect but it could make a significant difference if it is not dealt with properly.

To minimize the aforementioned unreliability, we utilize a mask to exclude the contribution from the X-ray rays passing through the two unreliable portions of the reconstructed volume in computing the loss function in Eq.~\ref{eq:LLE}. This mask is generated in the following three steps:
\begin{enumerate}
    \item Determine the unreliable portions and generate a volume mask (binary, zero for normal and one for unreliable voxels);
    \item Forward project the volume mask to the projection domain;
    \item Threshold the projected results and generate the unreliable volume masks in the projection domain.
\end{enumerate}

There are several different approaches to determine unreliable portions. For example, we can determine reliability according to the Tam-Danielsson window, or manually select the slice range per noise and image quality and reserve some margins. In this study, we used the latter one.

\subsection{Incremental updating}
In our previous work~\cite{chen2018general}, the algorithm directly estimates the motion vectors from an updated reconstruction in each iteration using the same dense sampling grid. However, the error of the motion estimation is expected to gradually decease as the algorithm iterates, which is also what we observed in practice. Global searching absolute values in a large grid may be computationally wasteful in later iterations. Hence, we propose a more efficient local searching strategy on an incrementally refined grid for relative value; i.e., in each new iteration, we predict correction to the motion estimation from the last iteration. Since the error is decreasing through iterations, we gradually shrink the sampling space while maintaining the same number of samples to generate a finer grid for improved searching accuracy. Furthermore, we use a much smaller number of samples compared to the absolute global searching to boost the searching efficiency without sacrificing performance.

\subsection{Virtual bed removal}
In most CT scans, an object to be reconstructed is supported by a bed or a holder, which is static in reference to the scanning geometry. For example, a couch in a medical CT scanner or a sample holder in a micro-CT system. In the reconstructed image, the stationary bed or holder shows its sharp edges in a good image quality, regardless any object motion. However, during the aforementioned motion correction process, those static structures are misaligned in reference to the object. In other words, after we compensate for the motion of the object, the bed or holder will be blurred and degraded in the subsequently reconstructed image. This will prevent the loss function in Eq.~\ref{eq:LLE} from being correctly minimized.

To avoid this undesired counterbalance effect, we virtually remove the bed or holder from projections, use the cleaned projections to perform motion correction, and then combined the reconstructed object with the previously reconstructed bed and holder if needed. The steps for removal of the bed or holder include:
\begin{enumerate}
    \item Reconstruct an image volume in the initial scan geometry;
    \item Segment the bed (i.e., all static portions) from the image volume with a proper margin and generate a bed mask;
    \item Set voxels outside the bed mask to zero in the reconstruction for a bed-only image volume;
    \item Forward project the bed-only volume to obtain the bed-only projections;
    \item Subtract the bed-only projections from the original projections to obtain the cleaned projections, with bed removed for motion estimation.
\end{enumerate}

\section{Results}
\subsection{Numerical simulation}
\begin{figure}[!t]
    \centering
    \includegraphics[width=\linewidth]{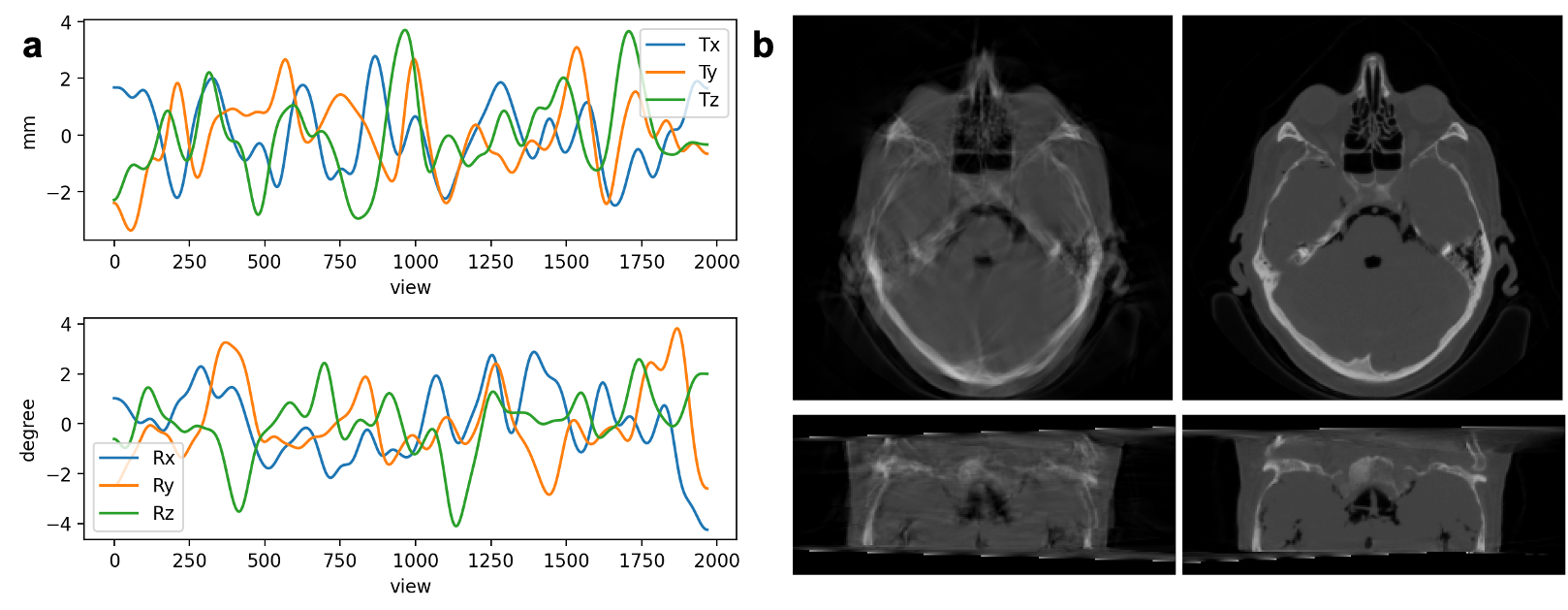}
    \caption{Motion parameters associated with the six degrees of freedom (a) and the axial and sagittal views of the corresponding reconstructions without motion compensation against the perfectly compensated reference (b), displayed in the window [0, 0.8] and unit of $cm^{-1}$.}
    \label{fig:MotionSim}
\end{figure}

First, the effects of unreliable volume masking and incremental updating were investigated in numerical simulation. The additional head image volume data from the visible human project~\cite{ackerman1998visible} was used as a digital phantom, and re-sampled into isotropic voxels of $0.5^3 mm^3$ in size. The detector was of 64 rows by 640 columns with an element area of $1mm\times 1mm$, and the scanning pitch is 43.13 $mm$ with 984 projections per rotation. To test the searching ability of our proposed method, we did not add noise initially. Two rotations were done with motions incorporated with six degrees of freedom, as shown in Fig.~\ref{fig:MotionSim} (a) where Ti and Ri denote the translation and rotation along and around the $i$th axis respectively. These motions caused severe artifacts in the reconstructed image, as demonstrated in Fig.~\ref{fig:MotionSim} (b).

\begin{figure}[!t]
    \centering
    \includegraphics[width=\linewidth]{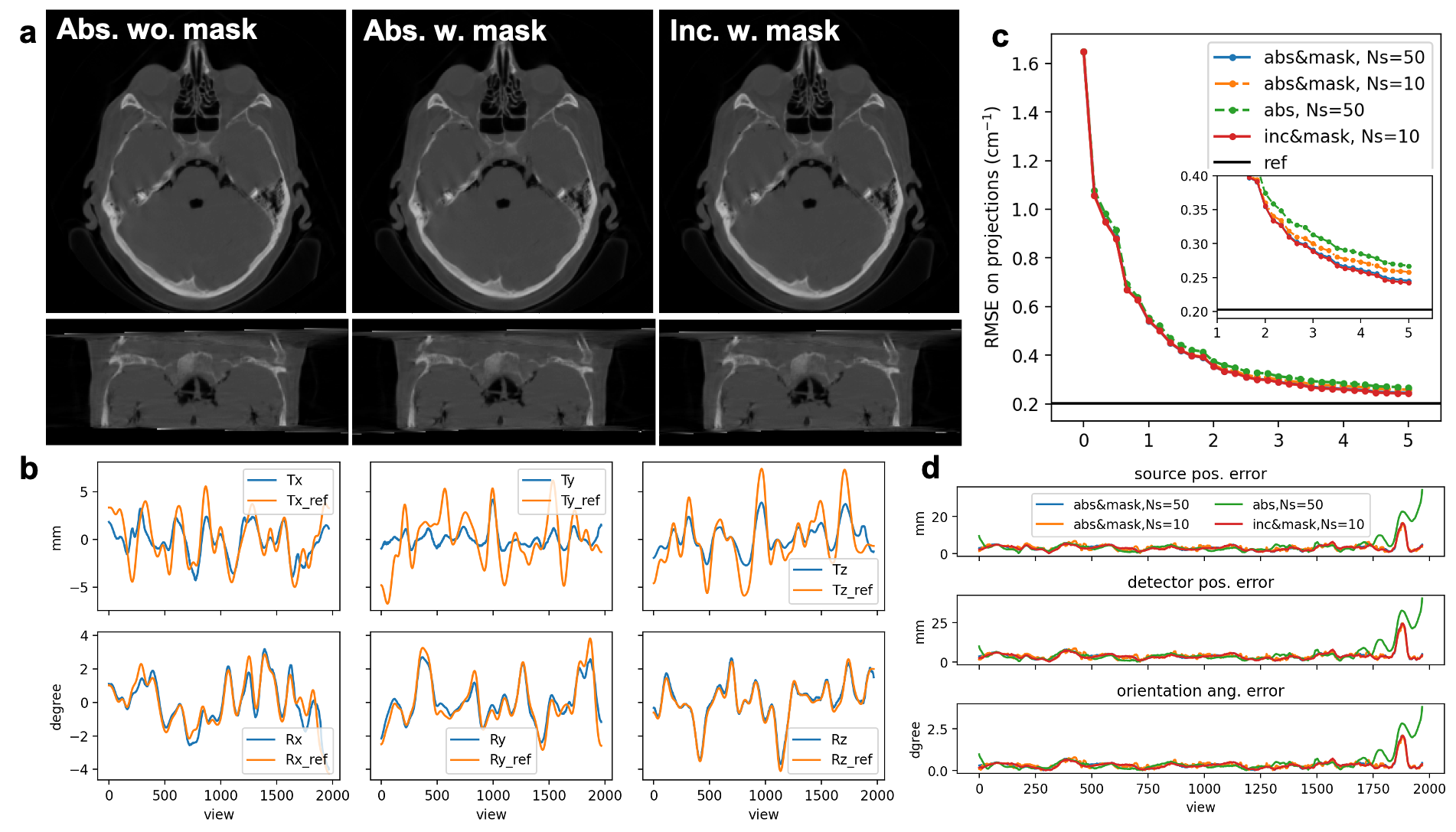}
    \caption{Numerical results of our motion estimation and correction in the four settings, including (1) absolute updating, without unreliable volume masking, $N_s = 50$ (abs. wo. mask); (2) absolute updating, with masking, $N_s = 50$ (abs. w. mask); (3) absolute updating, with masking, $N_s = 10$ (abs. w. mask); and (4) incremental updating, with masking, $N_s = 10$ (inc. w. mask). (a) The axial and sagittal views of compensated reconstructions in the settings 1, 2 and 4; (b) the estimated motion curves against the ground true with respect to the view, obtained in the 2nd setting; (c) the RMSE curves of the re-projections from the reconstructed image against the measurements during the motion correction iterations; and (d) the positional errors of the virtual source and detector positions and the angular error of the virtual detector normal against the ground truth.}
    \label{fig:SimResults}
\end{figure}

For motion estimation, each intermediate image was reconstructed as a volume of $240\times 240\times 100$ with a voxel size of $1^3 mm^3$. Figure~\ref{fig:SimResults}(a) shows the motion correction results using the LLE-based correction method. Those images were greatly improved relative to the original one. Despite remaining artifacts, they all came much closer to the ground truth that is the reconstructed images with perfect motion compensation, as shown in Fig.~\ref{fig:MotionSim}(b). Figure~\ref{fig:SimResults}(b) plots the motion curves along view, estimated with the absolute/fixed updating and unreliable volume masking, against the ground truth. The estimated curves align well with the references, especially for rotation motion, demonstrating the effectiveness of our LLE-based motion correction method.

The results obtained using different combinations of unreliable volume masking and incremental updating are compared in Figs.~\ref{fig:SimResults}(c) and~\ref{fig:SimResults}(d), including (absolute updating, masking, $N_s = 50$), (absolute updating, no masking, $N_s = 50$), (absolute updating, masking, $N_s = 10$), and (incremental updating, masking, $N_s = 10$), where $N_s$ denotes the size of the sampling grid, and masking means unreliable volume masking. Figure~\ref{fig:SimResults}(c) shows the change in the root mean squared error (RMSE) of the forward projections of the reconstructed image against the measurements with the iterative updating. Note that even with the perfect motion estimation the RMSE could not reach zero due to the large voxel size used for image reconstruction, and this RMSE is marked as the horizontal reference line in the figure. In all the four settings, the algorithm converged after 5 iterations, and the incremental updating with masking method worked the best, although it used a five times smaller sampling grid than that in the second best setting, which is the setting with absolute updating, masking and $N_s=50$. 
The full size sampling grid made the absolute/fixed updating with masking scheme perform better than a smaller sampling grid,  
and the absolute updating without masking scheme performed the worst even if it used a full size grid for sampling. To compare the motion estimation accuracy among these methods, we calculated the virtual spatial positions of the source and detector and virtual normal vector of the detector array assuming a static patient. The positional and angular errors against the ground truth are plotted in Fig.~\ref{fig:SimResults}(d). The other three settings except the no masking one led to closely overlapped curves, while the no masking setting gave a  
a cupping effect, as predicted in Subsection~\ref{sec:masking}, i.e., significantly larger mis-alignments were involved at the starting and ending views (corresponding to two ends of the image volume). Those results support the effectiveness of our proposed techniques described in the preceding section.

\subsection{Physical experiment}
The scan of a patient wrist was performed on a MARS Spectral clinical scanner equipped with a PCD array of 14 chips, each of which contains $128\times 128$ pixels of an area of $110\times 110 \mu m^2$ per pixel. The helical scan covers a FOV of 120mm in diameter and 100mm in longitudinal extent, with 373 projections per rotation and 5,713 projections in total. The source was operated at 120 kVp, 310 $\mu A$ with 0.25~mm Brass filtration. The source-to-detector distance and source-to-isocenter distance were 949~mm and 625~mm respectively. We selected one portion from the dataset of projections for demonstration, consisting of 746 projections (2 rotations, $128\times1792\times746$) to test our motion correction method. To minimize the noise influence, the channel with the lowest threshold (i.e., counting all photons of energy above 7keV) is selected out of 5 effective channels in total for motion estimation. The image was reconstructed into isotropic voxels of $180^3 \mu m^3$ per voxel, resulting a volume of $746\times 746\times 123$ voxels. The number of samples was set to 50 for each degree of freedom, and the range for translations in the x, y and z axes and the range for rotations around the x, y and z axes were defined as [-0.9 mm, 0.9 mm] and $[-2^\circ, 2^\circ]$ respectively. The SART algorithm was coded for reconstruction in 200 iterations. To optimize each of the six motion parameters, two iterations were performed, and the number of sequential updates was set to 5 for all the six freedoms (the main loop).

\begin{figure}[!t]
    \centering
    \includegraphics[width=\linewidth]{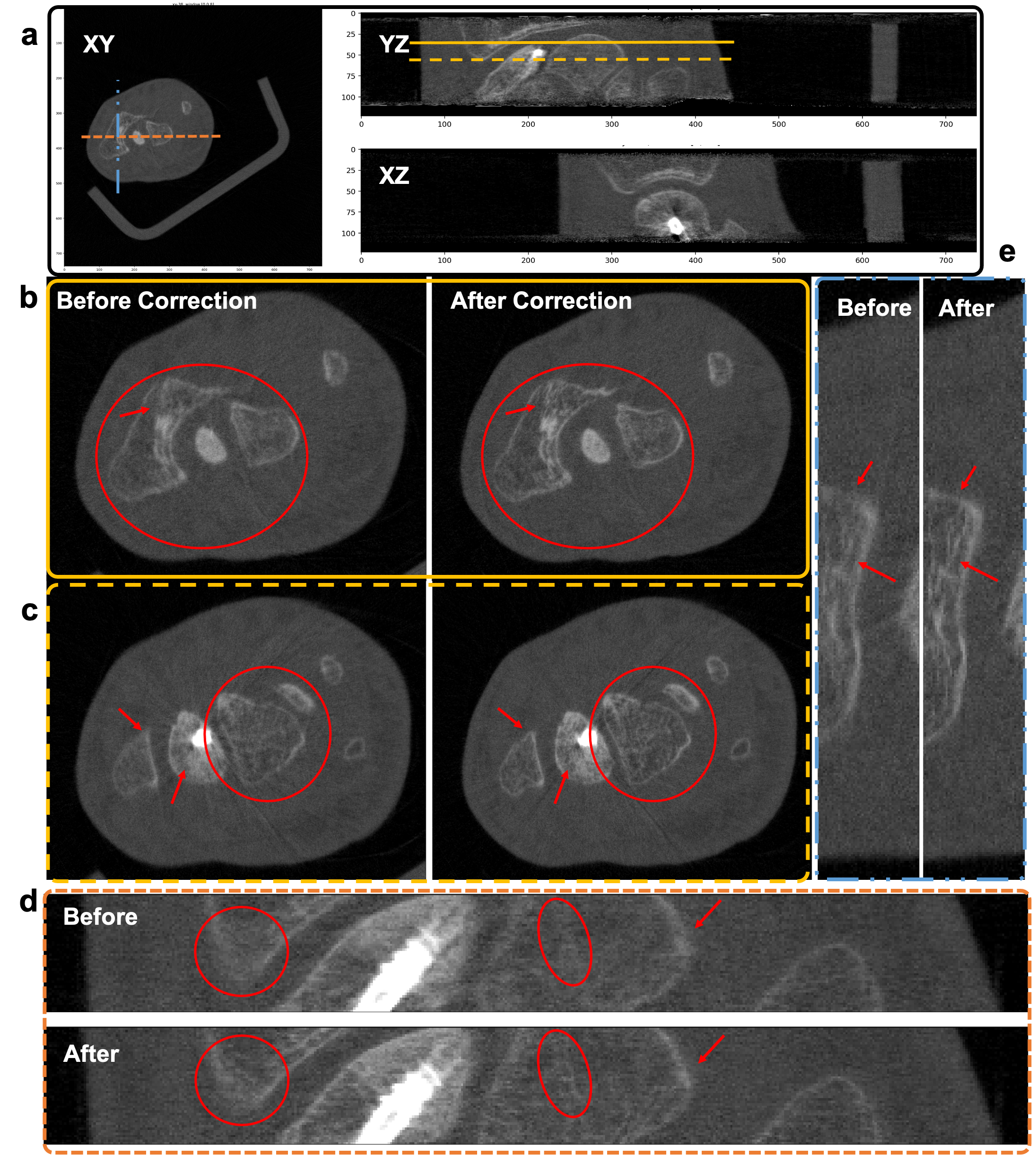}
    \caption{Motion correction for a real photon-counting CT scan of a living patient's wrist. (a) The reconstructed image volume before correction; (b) and (c) different axial slices before and after correction (the axial positions indicated in (a)); (d) and (e) are the coronal view and sagittal views in comparison (the cross-sectional positions marked in (a)). All the images are displayed in the window [0, 0.8] with the unit of $cm^{-1}$.}
    \label{fig:RealResult}
\end{figure}

Figure~\ref{fig:RealResult} shows the results before and after our motion correction. In Fig.~\ref{fig:RealResult} (b), the image is sharper (see the structures in the red circle) with subtle structures revealed from the blurry clouds after correction (see the region pointed by the red arrow). Similar comments can be made on Fig.~\ref{fig:RealResult} (c), where the blurry double-edge phenomenon was removed by the motion correction with fine features recovered in the bony region. Figure~\ref{fig:RealResult} (e) also demonstrates improved image resolution. In Fig.~\ref{fig:RealResult} (d), weak structures were enhanced in the red circles. Note that the missing chunk after reconstruction at the bottom right corner in Figs.~\ref{fig:RealResult} (b) and (c) is the bed that was virtually removed.

\begin{figure}[!t]
    \centering
    \includegraphics[width=0.8\linewidth]{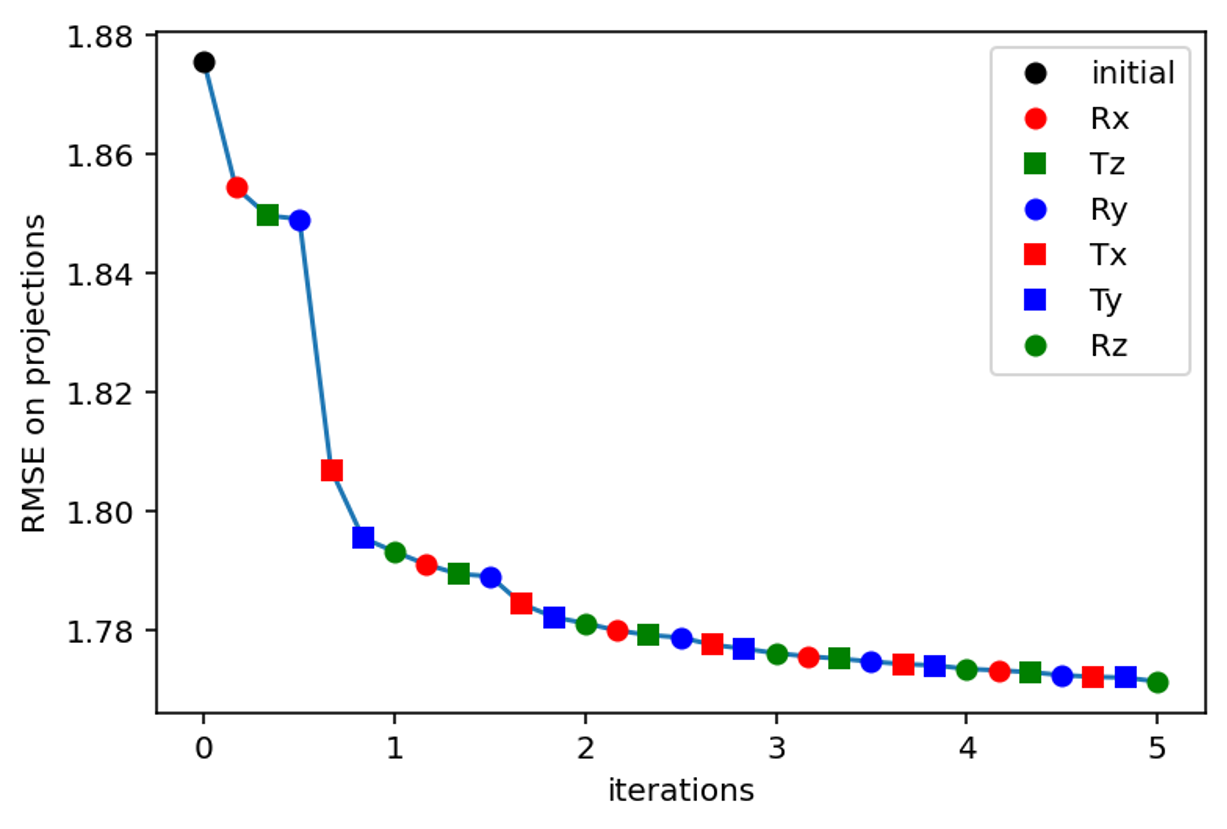}
    \caption{RMSE plots for projection data during the iterations for motion correction.}
    \label{fig:RealRMSE}
\end{figure}

Figure~\ref{fig:RealRMSE} plots the RMSE of the re-projections of the reconstructed volume against the measured projections with the bed removed through iterations. The six geometrical parameters was sequentially corrected.
The RMSE dropped rapidly after the first two iterations and then converged gradually, demonstrating the convergence and effectiveness of our method. Usually, three iterations were enough, and no significant structural difference was observed between the result with three iterations and that after five iterations.

\section{Conclusion}

In conclusion, we have presented an LLE-based motion correction method for helical photon-counting CT, which decomposes the motion correction problem into each and every views with respective to individual parameters, and works iteratively in a highly parallel manner. Our approach excludes bad photon-counting detector pixels, utilizes unreliable volume masking, incremental updating  and incrementally refined gridding techniques synergistically. Major improvements have been made in accuracy and efficiency of motion estimation and correction, as demonstrated in our numerical simulation and clinical experiments. Most importantly, our clinical study shows that significant resolution and contrast enhancement can be achieved in the high resolution regime, revealing subtle bony structures hidden by artifacts using our proposed LLE-based motion correction approach.

\section*{Acknowledgment}
This work is supported by fund from the National Institute of Health [National Cancer Institute (NCI)] under Award R01CA237267.

The authors would like to thank Dr. Mianyi Chen for helpful discussions and thank Niels De Ruiter for the help with reading the geometry of the raw projection data.

\ifCLASSOPTIONcaptionsoff
    \newpage
\fi

\bibliographystyle{IEEEtran}
%\bibliography{IEEEabrv,Mybibfile}

\end{document}